\documentclass[manuscript]{aastex}
\usepackage{emulateapj5}
\usepackage{apjfonts}
\usepackage{epsfig}
\usepackage{lscape}

\journalinfo{The Astrophysical Journal Letters in press}
\slugcomment{Received 2008 March 9; accepted 2008 June 9}
\shorttitle{Radiative Feedback in Red Quasars}
\shortauthors{WANG}

\def\Av{A_{\rm V}}
\def\cala{{\cal A}}

\def\calcr{{\cal C}_{\rm R}}
\def\calm{{\cal M}}

\def\calnc{{\cal N}_c}

\def\calq{{\cal Q}}

\def\nh{N_{\rm H}}

\def\pp{\prime\prime}

\def\ergs{\ifmmode {\rm erg~ s^{-1}} \else {\rm erg~s^{-1}}\ \fi}
\def\kms{\ifmmode {\rm km~ s^{-1}} \else {\rm km~s^{-1}}\ \fi}

\def\mbh{M_{\bullet}}

\def\mc{\multicolumn}
\def\mgii{\ifmmode Mg {\sc ii} \else Mg {\sc ii}\ \fi}

\def\pp{\prime\prime}

\def\sunm{M_{\odot}}

\def\lax{{$\mathrel{\hbox{\rlap{\hbox{\lower4pt\hbox{$\sim$}}}\hbox{$<$}}}$}}
\def\gax{{$\mathrel{\hbox{\rlap{\hbox{\lower4pt\hbox{$\sim$}}}\hbox{$>$}}}$}}

\def\ergs{${\rm erg~s^{-1}}$}

\begin{document}

\title{Extended multiwavelength fuzz around red quasars: observational appearance of
radiative feedback in action}

\author{Jian-Min Wang\altaffilmark{1,2}}

\altaffiltext{1}
{Key Laboratory for Particle Astrophysics, Institute of High Energy Physics,
Chinese Academy of Sciences, 19B Yuquan Road, Beijing 100049, China}

\altaffiltext{2}
{Theoretical Physics Center for Science Facilities, Chinese Academy of Sciences}

\begin{abstract}
Red quasars are a population, characterized by significant extinction in UV, which could be explained by 
absorption of dusty gas on a scale of a few kpc. We show that the enhanced radiation-pressure drives the 
dusty gas to supersonically expand and produces shocks. The shocks energize electrons to be relativistic 
via the first Fermi acceleration. As a balance of shock acceleration and synchrotron emission and 
inverse Compton scattering, the maximum Lorentz factor of the electrons reaches as $\sim 10^6$. The 
shocked interstellar medium appears as extended multiwavelength fuzz, in which synchrotron 
emission from the electrons peaks at near infrared or UV bands and inverse Compton scattering around 
1.0GeV$-$0.1TeV. Future multiwavelength images of the fuzz would provide new clues to study the details 
of radiative feedback if red quasars could be a certain phase in evolutionary chains of galaxies.
\end{abstract}

\section{Introduction}
Radiative feedback from quasars to surroundings of their hosts has been realized for many years.
Several effects have been suggested, including the Compton heating (Begelman 1985; Chang et al. 1987; 
Wang et al. 2005), acoustic wave heating (Chelouche 2007), proximity effects (Bajtlik et al. 1988; 
Goncalves et al. 2008; Wild et al. 2008; Prochaska \& Hennawi 2008). The strong feedback due to radiation 
may quench the inflows fueling the central black hole 
(e.g. Binney \& Tabor 1995) and star formation so as to arise the relation between black hole 
masses and dispersion velocities (Silk \& Rees 1998; King 2003; Ciotti \& Ostriker 2007; Fabian et al. 
2006). However, the direct large scale evidence for feedback is scarce, and even absent in most cases.

Traditionally quasars are selected based on their bluer color, 
however, the existence of significant population of red quasars has been the subject of debate for a number of 
years. The percentage of red quasars is only $\sim 6\%$ or so from the large database of Sloan Digital Sky 
Survey (SDSS) (Richards et al. 2003), $\sim 20-30\%$ in the FIRST-2MASS red quasar survey (Glikman et al. 2007),
and $\sim 20-30\%$ in {\em Spitzer} survey (Brown et al. 2006; Lacy et al. 2007). The fraction of red quasars 
remains a big debate, however this might be due to heterogeneous definition or selection effects of surveys. 
There is increasing evidence for dust extinction as one of the major explanations for red quasars
from SDSS (Richards et al. 2003), {\em GALEX} data (Trammell et al. 2007), {\em Spitzer} survey
(Brown et al. 2006; Lacy et al. 2007) and {\em HST} images (Urrutia et al. 2008) as well as from numerical
simulations (Di Matteo et al. 2005; Hopkins et al. 2007). The debris, ejected from mergers or starburst triggering 
the central black holes, surrounds the nucleus prior to the galactic winds (Sanders et al. 1989; Yun et al. 2004;
Fabian et al. 2006 and Hopkins et al. 2007), reddening quasars. The well-known roles of dust particles in outflows, 
as results of strong coupling between the charged grains with gas (e.g. Draine \& Salpeter 1979),
have been realized in the dynamics of dusty gas (Chang et al. 1987; Murray et al. 2005; Fabian et al. 2006).
It is then expected that the swept medium by quasar radiation results in extinction of quasars.

The present Letter makes an attempt to connect the action of the radiative feedback to the phenomena of red 
quasars and predicts some observational features. We argue that radiation from quasar is partly converted to 
kinetic luminosity of supersonic outflowing gas and then partly tansformed to the kinetic energy of accelerated 
electrons via Fermi mechanism, producing extended fuzz with non-thermal emission from radio to $\gamma-$rays around 
red quasars.

\section{Dynamics of dusty gas}
Dusty winds in homogeneous medium have been discussed 
in the literatures (Chang et al. 1988; Murray et al. 2005; Fabian et al. 2006). On the other hand, the gas in dark 
matter halos is driven into two phases in temperatures by the thermal instability (Fall \& Rees 1985), especially in
the presence of radiation heating of quasars. Thermal states of the medium are not the main goals of this Letter, 
but we will discuss two cases of the medium: 1) homogeneous interstellar medium (ISM) 
and 2) clumpy ISM, respectively. Observational appearance of the extended fuzz is quite similar for two cases 
unless the high spatial resolution of radio telescopes could resolve the cold clouds.

\subsection{Homogeneous ISM}
For an isothermal sphere of gas with density profile as $\rho\propto R^{-2}$, Murray et al. (2005) show 
the velocity of the momentum-driven winds at $R$ for the optically thin limit
\begin{equation}
V(R)=2\sigma \left[\frac{L}{L_{\rm Edd}}\left(1-\frac{R_{\rm exp}}{R}\right)+
     \ln \left(\frac{R_{\rm exp}}{R}\right)\right]^{1/2},
\end{equation}
from eq. (26) in Murray et al. (2005), where $R$ is the radius of the expanding winds,
$R_{\rm exp}=\sigma^2\sigma_{\rm T}/2\pi G m_p\approx 123~\sigma_{200}^2$pc is the initial radius of the winds 
(Fabian et al. 2006), $\sigma$ is the dispersion velocity ($\sigma=200\sigma_{200}$~km~s$^{-1}$), 
$L_{\rm Edd}=2.07\times 10^{46}\sigma_{200}^4$ergs~s$^{-1}$ 
is the Eddington luminosity, and $\sigma_{\rm T}$ is the Thompson cross section.

Here we only consider graphite grains here and assume they are sphere with
a radius $a$. We assume the ISM has the same gas-to-dust ratio with the Galaxy.
Since the charged particles of dust are strongly coupled with the ionized atoms via Coulomb interaction
(Draine \& Salpeter 1979), the dusty gas feels a stronger radiation pressure, amplified by a factor of 
\begin{equation}
\cala=\frac{\sigma_{\rm d}}{\sigma_{\rm T}}=4.7\times 10^2a_{-5}^2,
\end{equation}
where $\sigma_{\rm d}=3.1\times 10^{-22}a_{-5}^2~{\rm cm^{-2}}$ is the mean cross section per hydrogen atom 
for UV photons in the presence of dust (Martin \& Ferland 1980; Chang et al. 1988; Fabian et al. 2006) 
and $a_{-5}=a/10^{-5}~{\rm cm}$. The factor $\cala$ has been discussed for various cases in detail by Fabian et al. 
(2006). For dusty gas, the Eddington limit in eq. (1) should be replaced by $L_{\rm Edd}^{\prime}=L_{\rm Edd}/{\cal A}$.
For a quasar with a bolometric luminosity $L_{\rm Bol}$, the UV luminosity is approximate to 
$L_{\rm UV}=\xi L_{\rm Bol}=\xi \lambda L_{\rm Edd}$, where $\xi\approx 0.1$ (Marconi et al. 2004)
and $\lambda$ is the Eddington ratio. Inserting $L_{\rm Edd}^{\prime}$ into eq. (1), we have the maximum 
velocity of the expansion 
\begin{equation}
V_{\rm max}=2\sigma\sqrt{{\cal A}\lambda\xi-\ln \left({\cal A}\lambda\xi\right)-1}
           \approx 13.4\sigma  {\cal A}_{500}^{1/2}\left(\lambda\xi_{0.1}\right)^{1/2},
\end{equation}
at the radius
\begin{equation}
R_{_{V_{\rm max}}}={\cal A}\lambda\xi R_{\rm exp}\approx 6.0~{\cal A}_{500}\lambda\xi_{0.1}\sigma_{200}^2~{\rm kpc},
\end{equation}
and the Mach number is 
\begin{equation}
\calm_{\rm max}=V_{\rm max}/c_s=93.4~\cala_{500}^{1/2}\left(\lambda\xi_{0.1}\right)^{1/2}T_5^{-1/2},
\end{equation}
where $\cala_{500}=\cala/500$, $\xi_{0.1}=\xi/0.1$, $c_s=2.87\times 10^6T_5^{1/2}$cm~s$^{-1}$ is the sound speed and 
$T_5=T/10^5$K. Here we assume warm winds with a temperature of $10^5$K. The supersonic winds form shocks inevitably
and the strongest shocks are at $R_{_{V_{\rm max}}}\approx 6.0$kpc. It should be noted that the velocity from eq. (1) 
is insensitive to the radius after $R_{_{V_{\rm max}}}$, implying that strong shocks still extend further. Chang et al. 
(1987) showed that shocks appear in both one- and two-dimension flow by detailed dynamical calculations, which agrees 
with the present simple estimation. As a result of so strong shocks, electrons will be accelerated 
by the first order Fermi process (Blandford \& Eichler 1987).

Given the extinction coefficient $\Av$, the fraction of radiation absorbed by the dusty gas is 
$\alpha=1-10^{-0.4\Av}$. We assume that the absorbed energies are converted into the kinetic luminosity of the 
winds $L_{\rm kin}=\eta\alpha\xi\lambda L_{\rm Edd}
=1.8\times 10^{44}\eta_{0.5}\alpha_{0.17}\lambda\xi_{0.1}\sigma_{200}^4~{\rm ergs~s^{-1}}$,
where $\alpha_{0.17}=\alpha/0.17$ for $\Av=0.2$ and $\eta_{0.5}=\eta/0.5$ is the processing efficiency. 
The total energy channeled into relativistic electrons is given by $L_{\rm non-th}=f_aL_{\rm kin}$,
where $f_a$ is the acceleration efficiency converting the kinetic energy into the electrons.
Though the detailed process of the acceleration is still insufficiently understood, it is generally taken as 
$f_a=0.05$ (Blandford \& Eichler 1987). We then have the non-thermal luminosity from the electrons 
\begin{equation}
L_{\rm non-th}=0.9\times 10^{43}f_{0.05}^{}\eta_{0.5}\alpha_{0.17}\lambda\xi_{0.1}\sigma_{200}^4~{\rm ergs~s^{-1}},
\end{equation}
where $f_{0.05}=f_a/0.05$. The spectrum emitted from the electrons will be given in \S3.

\subsection{Clumpy ISM}
The physics of the infalling gas has been extensively discussed stemming from Fall \& Rees (1985). The basic
constraints on the medium are from the gravitational collapse and thermal conduction between the hot and cold phases
(Fall \& Rees 1985; Mo \& Miralda-Escude 1996). The typical temperature and density of the hot gas are of $T_h=10^6$K 
and $n_h=10^{-2}$cm$^{-3}$ in elliptical galaxies, respectively (Mathews \& Brighenti 2003).

The lower limit of cold clouds is given by $M_{\rm cond}=5.0\times 10^4R_{6\rm kpc}^2\sunm$ due to thermal conduction, 
otherwise they will be evaporated (Fall \& Rees 1985), whereas
the clouds will collapse into smaller ones if they are too massive (e.g $\sim 10^6\sunm$). 
We assume the clouds are sphere with radius $R_c$ and mass $M_{\rm c}=qM_{\rm cond}$ ($q\ge 1$). A pressure balance 
$n_cT_c=n_hT_h$ holds between the clouds and their surroundings. The geometric covering factor can be defined as
\begin{equation}
\calcr=\left(\frac{\Delta \Omega}{4\pi}\right)\calnc,
\end{equation}
where $\Delta \Omega$ is the solid angle of an individual cloud and $\calnc$ is the total number of the cold clouds.
This covering factor is the same with the probability that a quasar is a red one.
Here we assume that {\em all} the clouds are identical with same size, shape and physical conditions. For simplicity,
we assume that the clouds are sphere. If the radius of the clouds is $R_c$ and $R_c\ll R$, we have 
$\Delta \Omega/4\pi =R_c^2/2R^2$, where $R$ is the distance of the clouds to the galactic center. 
The number of the clouds is then estimated from eq. (7) if the geometric covering factor is estimated from the
percentage of red quasars.  The individual cloud should not be too dense, otherwise the red quasars become type 2s. 
This constraint can be set up by the column density $\nh$ through the extinction coefficient $\Av$.
We then have the density and temperature
\begin{equation}
n_c=8.4~q^{-1/2}N_{21}^{3/2}~{\rm cm^{-3}},
\end{equation}
\begin{equation}
T_c=1.2\times 10^3~q^{1/2}n_{h,-2}T_{h,6}~{\rm K}, 
\end{equation}
the radius $R_c=38.7~q^{1/2}N_{21}^{-1/2}~{\rm pc}$ and the cloud number is
$\calnc=1.3\times 10^3~{\cal C}_{\rm R0.1}q^{-1}N_{21}R_{\rm 6kpc}^2,$
where ${\cal C}_{\rm R0.1}=\calcr/0.1$, $T_{h,6}=T_h/10^6$K and $n_{h,-2}=n_h/10^{-2}~{\rm cm^{-3}}$. 
Though we do not know the details of the parameter $q$, we regard eq. (8-9) as the typical values for 
$n_c$ and $T_c$ since they are not very sensitive to $q$. The Mach number of the dusty winds in 
the clouds is 
\begin{equation}
\calm=9.3\times 10^2~{\cal A}_{500}^{1/2}\left(\lambda\xi_{0.1}\right)^{1/2}T_3^{-1/2},
\end{equation}
where $T_3=T_c/10^3$K. The Mach number is stronger than the homogeneous ISM since the temperature is lower in the
clouds. The shocked clouds produce relativistic electrons. 

When a cloud is just on the line of an observer's sight, the quasar appears as one red quasar.
The transparent luminosity through the clouds is given by $L_1=\epsilon \calnc \left(\Delta\Omega/4\pi\right)L_{\rm in}$,
where $\epsilon=10^{-0.4A_{\rm V}}$ and $L_{\rm in}$ is the intrinsic luminosity of the quasar, while the unabsorbed
luminosity is given by $L_2=(1-\calcr)L_{\rm in}$. The global energy budget is then given by $L_{\rm obs}=L_1+L_2$, we 
have $L_{\rm obs}/L_{\rm in}=\epsilon \calcr+(1-\calcr)$ and $1-L_{\rm obs}/L_{\rm in}=(1-\epsilon)\calcr$ means the 
fraction of the intrinsic luminosity converted into the kinetic energy of the winds. The kinetic
luminosity of the winds is then given by $L_{\rm kin}=(1-\epsilon)\calcr \eta\xi\lambda L_{\rm Edd}$.
The total non-thermal luminosity from the relativistic electrons is given by
\begin{equation}
L_{\rm non-th}=f_a(1-\epsilon)\calcr \eta \xi\lambda L_{\rm Edd}.
\end{equation}
Eq.(1) is still valid for the clumpy ISM (Murray et al. 2005), but the shocks are produced mainly on the cold clouds. 

\section{Electron Acceleration and emission}
Some electrons will be energized to be relativistic through Fermi shock acceleration. The maximum energy of the 
relativistic electrons is determined by the energy gain and loss. Here we focus on the non-thermal radiation
from the relativistic electron, rather than the detail of acceleration process. 
In this Letter, we assume the synchrotron radiation and inverse Compton scatter are two main non-thermal radiation 
processes. Since the magnetic field  is poorly understood in such 
a context, we assume an equipartition between the tangled magnetic field 
and the gas, then the energy density of the magnetic field $U_{\rm B}=1.38\times 10^{-12}(n_cT_c)_4~$ergs~cm$^{-3}$
and $B=5.9~\left(n_cT_c\right)_4^{1/2}\mu$G, where $(n_cT_c)_4=n_cT_c/10^4{\rm K~cm^{-3}}$. 
There are three sources of the seed photons of Compton scattering: 1) from the
nuclear of quasars; 2) the reprocessed photons by the clouds; 3) synchrotron photons 
and 4) from the host galaxies. However, light from hosts can be neglected for quasars bright enough. The second 
could be neglected since they are only $\sim 10\%$ of the first source 
(if the optical depth $\tau\approx \Av$ for $\Av=0.1$, but may be important
for the very red quasars or redden quasars).  The energy loss is mainly due to synchrotron radiation and inverse Compton 
scattering of the seed photons from quasar's nuclear. 

The energy density of quasar's radiation is given by
$U_{\rm qso}=7.7\times 10^{-11}L_{46}R_{6\rm kpc}^2~{\rm ergs~cm^{-3}}$ at $R$ from the center, 
where $R_{\rm 6kpc}=R/{\rm 6kpc}$ and $L_{46}=L/10^{46}$ergs~s$^{-1}$. We find that $U_{\rm qso}\gg U_{\rm B}$, 
indicating that the energy loss is mainly due to external inverse Compton scattering provided by the red quasars. 
The synchrotron self-Compton scattering can be neglected in such a context. For simplicity, we assume the quasar 
spectrum is characterized by a black body with a temperature of $10^4$K 
from the standard accretion disk (Inoue \& Takahara 1996). The timescale of energy loss due to 
the inverse Compton scattering is $t_{\rm loss}=3.0\times 10^7 \gamma^{-1}U^{-1}$ sec, where $U\approx U_{\rm qso}$,
whereas the acceleration time scale is $t_{\rm acc}=R_{\rm L}c/v_{\rm sh}^2$ (Blandford \& Eichler 1987), where 
$R_{\rm L}$ is the Lamore radius, and  $v_{\rm sh}\approx V_{\rm max}\approx 10^8$cm~s$^{-1}$ is the shock velocity.
The maximum Lorentz factor of the electrons is given by $t_{\rm loss}=t_{\rm acc}$,
\begin{equation}
\gamma_{\rm max}=7.7\times 10^6~v_{\rm sh,8}U_{-10}^{-1/2}B_{-6}^{1/2},
\end{equation}
where $U_{-10}=U_{\rm qso}/10^{-10}{\rm ergs~cm^{-3}}$, $v_{\rm sh,8}=v_{\rm sh}/10^8{\rm cm~s^{-1}}$ 
and $B_{-6}=B/10^{-6}{\rm G}$. The synchrotron radiation gets a peak at
\begin{equation}
\nu_{\rm syn}=4.5\times 10^{13}B_{-6}\gamma_{\rm max,7}^2~{\rm Hz},
\end{equation}
where $\gamma_{\rm max,7}=\gamma_{\rm max}/10^7$. The peak frequency due to inverse Compton scattering is
\begin{equation}
\nu_{\rm IC}\approx \gamma^2\nu_{\rm disk}=3.6\times 10^{27}\left(\gamma/\gamma_{\rm max}\right)^2T_4~{\rm Hz},
\end{equation}
where $\nu_{\rm disk}=3kT_{\rm disk}/h$,  $h$ is the Planck constant and $T_{\rm disk}=10^4T_4$K is the 
maximum temperature in Shakura$-$Sunyaev disk for quasars.  We assume a simple power law of the electrons
as $N(\gamma)=N_0\gamma^{-n}$ in a range of $\gamma_{\rm min}\le\gamma\le \gamma_{\rm max}$
via the first order Fermi acceleration, where $\gamma_{\rm min}$ is the minimum Lorentz factor. We treat 
$\gamma_{\rm min}$ as a free parameter in this paper. $N_0$ is determined by eqs. (6) or (11). 
For a red quasar containing a black hole with $\mbh=10^8\sunm$, we plot the multiwavelength continuum of the
fuzz for three different $n$ in Fig 1. It should be noted that the difference between $\nu_{\rm IC}$ and the 
spectrum in Fig. 1 is due to the inclusion of the Klein-Nishina effects in the later.

Some uncertainties of the proposed model should be noted, which are most likely because of the dependence of the 
factor ${\cal A}$ on the thermal state of the medium as shown in Fabian et al. (2006). Here we use the maximum
${\cal A}$. The $\gamma_{\rm max}$, depending on the shock velocity, 
could be lowered when ${\cal A}$ decreases, leading to the synchrotron peak shifts to long wavelength as well as
the inverse Compton peak. We use a magnetic field in an equipartition with thermal energy of medium, but it can 
not be deviated from the equipartition too much.

\section{Detecting the fuzz}
The isothermal sphere under the galactic potential gives the density
$\rho(R)=f_g\sigma^2/2\pi GR^2$, we have the column density of the gas exterior to $R_{\rm exp}$,
$N_{\rm H}=f_g\sigma^2/2\pi Gm_pR=3.0\times 10^{21}f_{g0.1}\sigma_{200}^2R_{\rm 6kpc}^{-1}~{\rm cm^{-2}}$, where 
$f_g=0.1f_{g,0.1}$ is the gas fraction (Fabian et al. 2006). This leads to an extinction coefficient 
$\Av=0.2\calq_{500}^{-1}N_{21}$, appearing as a red quasar, where $N_{21}=N_{\rm H}/10^{21}{\rm cm^{-2}}$
and $\calq_{500}=\calq/500$ is the gas-to-dust 
ratio (Gorenstein 1975).  We note that the sweeping timescale of the radiation-driven winds is 
$\tau_{\rm w}=R/V(R)=2.2\times 10^6R_{\rm 6kpc}\left(V/V_{\rm max}\right)^{-1}$ yr.  The lifetime of quasars 
$\tau_{\rm qso}$ is about the Salpeter time $\tau_{\rm Salp}=4.5\times 10^7\eta_{-0.1}$ yr ($\eta=\eta/0.1$ is the
radiative efficiency), we have the percentage of red quasars at a level of $\tau_{\rm w}/\tau_{\rm qso}\sim 5.0\%$. 
This roughly agrees with the results from SDSS (Richards et al. 2003).

\figurenum{1}
\centerline{\psfig{figure=f1.ps,angle=-90,width=8.5cm}}
\vglue 0.2cm
\figcaption{\footnotesize
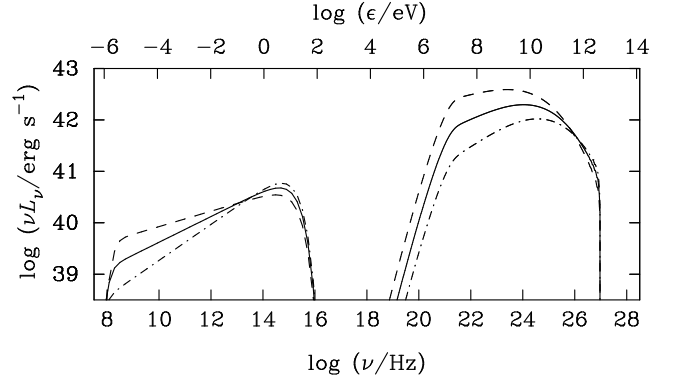
The multiwavelength continuum from radio to $\gamma-$ray bands. We take $L_{\rm non-th}=10^{43}~{\rm ergs~s^{-1}}$,
$\gamma_{\rm max}=7.7\times 10^6$, $\gamma_{\rm min}=900$, 
$U=10^{-10}~{\rm ergs~cm^{-3}}$  and $B=6.0\mu$G. The solid, dotted and dashed lines 
correspond to the indexes of $n=2.5$, $2.7$ and $2.3$, respectively. The Klein-Nishina effects are included in the 
inverse Compton scattering.
}
\label{fig1}
\vglue 0.5cm

\begin{center}
\centerline{\small \sc Table 1. Detection of the multiwavelength fuzz}
{\footnotesize
\begin{tabular}{lllllllll}\hline\hline
            &\mc{3}{c}{$z=0.01,~\theta\approx 30^{\pp}$} & &\mc{3}{c}{$z=0.2,~\theta\approx2^{\pp}$}\\ \cline{2-4}\cline{6-8}
Instruments & $F$  & $I$ & note    & & $F$ & $I$ & note\\
(1)         & (2)  & (3) & (4)     & & (5) & (6) & (7) \\ \hline
ALMA        &47.4  &16.8 & yes     & &0.09 &0.3  & yes \\
MIPS        &3.0   &1.0  & yes     & &.... &...  & no  \\ 
JWST        &0.4   &0.14 & yes     & &0.7  &2.6$\times 10^{-3}$&yes  \\
GLAST       &150   &...  & yes     & &0.3  &...  & yes \\
HESS (I)    &0.03  &...  & marginal& &...  &...  & no  \\
VERITAS     &0.03  &...  & marginal& &...  &...  & no  \\ 
HESS (II)   &0.03  &...  & yes     & &...  &...  & no  \\ \hline
\end{tabular}
\vskip 0.1cm
\parbox{3.45in}
{
\indent
Note: $\theta$ is the angular radius (in arcsec) at $z$. Col. (1) is the name of instrument; Col. (2-4) flux (in mJy),
brightness ($\mu$Jy~arcsec$^{-2}$) for ALMA, MIPS and JWST and note, respectively, but in $10^{-9}$ phs~s$^{-1}$~cm$^{-2}$ 
for $\gamma-$ray detectors; Col. (5-7) are the same with (2-4), but in $\mu$Jy, $\mu$Jy~arcsec$^{-2}$ for $z=0.2$.
"yes" stands for detection, "no" for non-detection and "marginal" marginal detection. We assume the cosmological constants 
$H_0=71~{\rm Mpc^{-1}~km~s^{-1}}$, $\Omega_{\rm M}=0.27$ and $\Omega_{\Lambda}=0.73$.\\
{\em ALMA}: Atacama Large Millimetre/Submillimetre Array, spatial resolution $\Delta \theta=0.005^{\prime\prime}$ at 
$\nu=900$GHz.\\
{\em MIPS}: Multiband Imaging Photometer for Spitzer, $\Delta\theta=5^{\prime\prime}$, working at  
24, 70 and $160\mu$m. Here is for 24$\mu$.\\ 
{\em JWST}: J. Webb Space Telescope. Here is for NIR camera at $5\mu$m.\\
{\em GLAST}: Gamma-ray Large Area Space Telescope ($>100$MeV), sensitivity $\sim 10^{-9}$ phs~s$^{-1}$~cm$^{-2}$.\\
{\em HESS}: High Energetic Spectroscopic System ($>100$GeV), sensitivity $\sim 10^{-10}$  
(Phase I) and $\sim 10^{-11}$ phs~s$^{-1}$~cm$^{-2}$ (Phase II).\\
{\em VERITAS}: Very Energetic Radiation Imaging Telescope Array System
}}
\end{center}

The targets for searching the fuzz can be found from those low redshift red quasars from FIRST-2MASS, {\em Spitzer} 
and Sloan samples. Table 1 gives the future detections of the fuzz through multiwavelength telescopes.
The predicted fuzz has a level of synchrotron power of $10^{40\sim 41}$ergs~s$^{-1}$, which is easily to be detected 
by radio telescopes. IR band telescopes are obviously capable of imaging the fuzz, in particular, the ALMA and JWST 
are able to detect the fuzz at $z>0.2$ in future. The fuzz could be marginally detected by the existing {\em HESS} 
and {\em VERITAS}, but it is worth making an attempt. {\em GLAST} is anticipated to explore fuzz more feasibly. 
We have to note that TeV emission from fuzz could be diluted via $\gamma-\gamma$ interaction
by the cosmic IR back ground photons beyond redshift $z=0.2$. So it is not expected for $\gamma-$ray telescopes
to detect the TeV emission from fuzz for red quasars with $z>0.2$. If the episodic activity of supermassive black 
holes is really caused by a series of mergers of galaxies (Marconi et al. 2004; Di Matteo et al. 2005;  Wang et al. 
2006; 2008), there is no doubt that the detection of the fuzz will provide the most powerful
diagnostic to evolutionary chains of galaxies and quasars from details of the radiative feedback.

Finally, it should be mentioned that X-ray emission from fuzz is extremely faint unless the maximum Lorentz factor 
is lowered by a factor of 10. This could be plausible if the factor ${\cal A}$ decreases by a factor of 10. In such 
a context, the fuzz is able to produce hard X-rays detected by the {\em Nustar} (The Nuclear Spectroscopic Telescope 
Array).

\section{Conclusions and discussions}
In this Letter, we show that the enhanced radiation pressure inevitably leads to strong shocks in the dusty gas
of both homogeneous and clumpy ISM. The first Fermi shock acceleration results in production of relativistic electrons
in the shocked ISM, from which produces non-thermal emission via synchrotron radiation and inverse Compton 
scattering, displaying multiwavelength fuzz from radio, infrared and $\gamma-$ray bands.
The current radio telescopes, MIPS, HESS and VERITAS are capable of detecting the fuzz.
The multiwavelength image of the fuzz provides an interesting and new clue to understand
the interaction between quasars and their environments.

The predicted $\gamma-$ray emission from fuzz could have significant contribution to cosmic $\gamma-$ray background
if {\em all} quasars have to undergo such a red phase during their cosmic evolution. We will issue this problem in 
the future.

\acknowledgements{I am grateful to the referee for a thoughtful report improving the manuscript.
I acknowledge Ari Laor for useful discussions on dynamics of dusty gas, V. Wild, L. C. Ho, 
S.-N. Zhang and X.-Y. Xia for interesting discussions. I appreciate the stimulating discussions among the members 
of IHEP AGN group, especially C. Hu and Y.-M. Chen for interesting conversations. The research is supported by NSFC 
and CAS via NSFC-10325313, 10733010 and 10521001, and KJCX2-YW-T03, respectively.}

\end{document}